\newcommand{\simgt}{\,\rlap{\lower 3.5 pt \hbox{$\mathchar \sim$}} \raise
1pt \hbox {$>$}\,}
\newcommand{\simlt}{\,\rlap{\lower 3.5 pt \hbox{$\mathchar \sim$}} \raise
1pt \hbox {$<$}\,}

\def\be{\begin{equation}}
\def\ee{\end{equation}}
\def\ba{\begin{eqnarray}}
\def\ea{\end{eqnarray}}

\def\dmeV{\delta m^2_{\rm eV}}

\def\dm{\delta m^2}
\def\dmeV{\delta m^2_{\rm eV}}

\def\l{\left}
\def\r{\right}
\def\ot{\frac{1}{2}}

\def\nn{\nonumber}

\def\MeV{{\,{\rm MeV}}}
\def\eV{{\,{\rm eV}}}

\def\ot{\frac{1}{2}}

\def\fas{f_{(a-s)}}
\def\lexpl{l_{\rm expl}}

\documentclass[12pt]{article}
\usepackage[dvips]{epsfig}
\begin{document}

\setlength{\baselineskip}{0.30in}

\newcommand{\nc}{\newcommand}
\newcommand{\num}{\nu_\mu}
\newcommand{\nue}{\nu_{\rm e}}
\newcommand{\nut}{\nu_\tau}
\newcommand{\nus}{\nu_{\rm s}}
\newcommand{\mnus}{m_{\nu_{\rm s}}}
\newcommand{\taus}{\tau_{\nu_{\rm s}}}
\newcommand{\nnt}{n_{\nu_\tau}}
\newcommand{\rnt}{\rho_{\nu_\tau}}
\newcommand{\mnt}{m_{\nu_\tau}}
\newcommand{\tnt}{\tau_{\nu_\tau}}
\newcommand{\bi}{\bibitem}
\newcommand{\rar}{\rightarrow}
\newcommand{\lar}{\leftarrow}
\newcommand{\lrar}{\leftrightarrow}
\newcommand{\so}{\, \mbox{\sin}\Omega}
\newcommand{\co}{\, \mbox{\cos}\Omega}
\newcommand{\sotil}{\, \mbox{\sin}\tilde\Omega}
\newcommand{\cotil}{\, \mbox{\cos}\tilde\Omega}
\makeatletter
\def\alt{\mathrel{\mathpalette\vereq<}}
\def\vereq#1#2{\lower3pt\vbox{\baselineskip1.5pt \lineskip1.5pt
\ialign{$\m@th#1\hfill##\hfil$\crcr#2\crcr\sim\crcr}}}
\def\agt{\mathrel{\mathpalette\vereq>}}

\newcommand{\eq}{{\rm eq}}
\newcommand{\tot}{{\rm tot}}
\newcommand{\M}{{\rm M}}
\newcommand{\coll}{{\rm coll}}
\newcommand{\ann}{{\rm ann}}
\makeatother

\renewcommand{\thefootnote}{\fnsymbol{footnote}}
\setcounter{footnote}{1}

\ \vspace*{-3.cm}
\begin{flushright}
  {MPI--PhT/2000-31}\\
  {\ }
\end{flushright}

\vspace*{0.3cm}

\begin{center}
\vglue .06in
{\Large \bf {\boldmath Lepton asymmetry creation in
the Early Universe.}}
\bigskip
\\{\bf R.~Buras \footnote{e-mail: {\tt rburas@mppmu.mpg.de}}} \\
{\it{Max-Planck-Institut f\"ur Physik (Werner-Heisenberg-Institut)\\
F\"ohringer Ring 6, 80805 M\"unchen, Germany
}}
\\{\bf D.V.~Semikoz \footnote{e-mail: {\tt semikoz@mppmu.mpg.de}}} \\
{\it{Max-Planck-Institut f\"ur Physik (Werner-Heisenberg-Institut)\\
F\"ohringer Ring 6, 80805 M\"unchen, Germany\\
and\\
Institute of Nuclear Research of the Russian Academy of Sciences\\
60th October Anniversary Prospect 7a, Moscow 117312, Russia}}
\\[.40in]
\end{center}

\begin{abstract}
Oscillations of active to sterile neutrinos with a small mixing angle
$\sin 2 \theta <10^{-2}$ could generate a large lepton asymmetry in
the Early Universe. The final order of magnitude of the lepton
asymmetry $\eta$ is mainly determined by its growth in the last stage
of evolution, the so called power-law regime. There exist two
contradictory results in the literature, $\eta \propto T^{-1}$ and
$\eta \propto T^{-4}$, where $T$ is the background medium
temperature. In the first case, the lepton asymmetry does not exceed
values of $10^{-4}$ for $|\delta m^2| \leq 1 \eV^2$, while in the second
case it can become larger than $10^{-1}$.  In this work we
analytically investigate the case $\eta \propto T^{-1}$, using a new
approach to solve the kinetic equations. We find that the power-law solution
$\eta \propto T^{-1}$ is not self-consistent. Instead, we find the
power law $\eta \propto T^{-11/3}$ to be a good approximation, which
leads to a large final asymmetry.

\end{abstract}

\newpage
\renewcommand{\thefootnote}{\arabic{footnote}}
\setcounter{footnote}{0}
\section{Introduction}

Neutrinos play an important role during Big Bang Nucleosynthesis
(BBN) due to their contribution to the total energy density.
Furthermore, the electron neutrinos directly influence the
neutron-proton reactions.  In this context, the Standard Model with
its three active massless neutrinos agrees very well with the
observations of the primordial light element abundances: a recent
analysis \cite{tytler0001318} of observational data claims that any
particles beyond the Standard Model must not contribute more than 20\%
to the total energy density of a single neutrino flavor.

The recent discovery of $\nu_\mu$-$\nu_\tau$ oscillations at the
Super-Kamiokande detector \cite{SK} has proven the existence of
neutrino masses. Reasonable neutrino masses of order $m_\nu \leq 1 \eV$
will not spoil the standard BBN picture, because BBN takes place at
temperatures $T \sim 0.1 \MeV\gg m_\nu$.  Also, mixing between
the active neutrino flavors plays no role since they are equally
populated during the relevant epoch.

But the situation dramatically changes if one supposes the existence
of one or more sterile neutrinos: even a partial excitation of more
than 20\% of a sterile neutrino state, e.g.~via oscillations, will
contradict the observational data of light element abundances.
Actually, introducing sterile neutrinos is quite natural, since
right-handed neutrinos would necessarily be sterile in the Standard
Model sense!

Another important place in which sterile neutrinos can play a role is
the creation of lepton asymmetry in the active neutrino sector. For
example, if a significant asymmetry between $\nu_{\rm e}$ and
$\overline{\nu}_{\rm e}$ is established, this could change the
neutron-proton ratio and lead to a contradiction with the
observational data.

Early considerations of the simplified kinetic equations found no
significant increase of the asymmetry in the case where the sterile
neutrino has a mass $m_{\nu_{\rm s}} \ll 1 \MeV$ and any vacuum mixing
angle and concluded that this asymmetry was always small, $\eta \ll 1$
\cite{enkvist90,barbieri91}.  However, later it was found that in the
region of small mixing angles the asymmetry can increase
significantly and reach values $\mathcal{O}(0.1)$ \cite{foot96}.
This statement was confirmed by numerical calculations \cite{fv96}
where a large final value of the asymmetry was found for the first time.
Recently, more accurate numerical calculations also confirmed the
increase of asymmetry to large values $\mathcal{O}(0.1)$
\cite{footAP99,bfoot00}. The reason for such a big increase of the asymmetry
is that it grows according to a power law close to $T^{-4}$.

Contrary to this statement is the result of an analytical approach by
Dolgov {\it et al.} \cite{dolgov99}, who found an additional
counterterm in the effective equation for the asymmetry, such that the
power law would change from $T^{-4}$ to $T^{-1}$. Thus the final
asymmetry would be $\mathcal{O}(10^{-4})$.

In this paper, we analyze the method of \cite{dolgov99} and show that
it is valid only when the evolution of the kinetic equations dictated
by the collision terms, i.e.~when the temperature of the primordial
plasma is large enough, $T \gg 1 \MeV$. For smaller temperatures, $T
\le$ \hbox{2-3} MeV, collisions become less important, and the
evolution of the neutrinos can be described by the well-known
Mikheyev-Smirnov-Wolfenstein (MSW) effect \cite{msw}. In this small
temperature region the approach of \cite{dolgov99} is not valid.
This effect has been studied in \cite{fv96}.

The counterterm found in \cite{dolgov99}, which leads to the power law
$T^{-1}$, only becomes compatible to the main term when the applied
perturbative expansion breaks down. This fact suggests that the
power law $T^{-1}$ could be wrong, but strictly speaking it does not
prove this. Therefore, we develop a new independent analytical method
which can be applied in the regions where the method of
\cite{dolgov99} does not work. Instead of expanding in the
coefficients of the kinetic equations, we solve these equations
pertubatively for the case when the creation rate of sterile neutrinos
is not significantly large.  Using this new method we show that the
$T^{-1}$ power law is not a self-consistent solution for the lepton
asymmetry. Instead, the power law $T^{-11/3}$ is approximately valid,
and therefore the asymmetry can reach large values of order $0.1$,
although our new method in its present form does not allow us to
calculate the final value of the asymmetry.

In section 2 we define the kinetic equations. Then in section 3
we shortly discuss the evolution of the lepton asymmetry before it
reaches the power-law regime and explain the power-law solution known
in literature. In section 4 we discuss some recent papers on this
topic, especially the one claiming the $T^{-1}$ power law
\cite{dolgov99}. In section 5 we present an improved approach similar to
\cite{dolgov99} and discuss its validity. In section 6 we present a new
analytic solution of the kinetic equations which we use to check the
validity of the $T^{-1}$ power law. In section 7 we conclude and
summarize our results. In the Appendix we comment on some numerical
fits.

\section{Kinetic equations}

In order to describe the evolution of the system of oscillating active
$\nu_{\rm a}$, ${\rm a} = {\rm e}, \mu, \tau$, and sterile neutrinos
$\nu_{\rm s}$ with a mass difference $\delta m^2$ and a vacuum mixing
angle $\theta$ in the Early Universe we need to consider the density
matrix formalism \cite{dolgov81,stodolsky87} and take into account
second order effects in the Fermi coupling constant
\cite{dolgov81,raffelt93, sigl93}.  Then the following set of
equations for the components of the density matrix can be found:
\ba
i(\partial_t -Hp\partial_p) \rho_{aa}
&=& F_0(\rho_{sa}-\rho_{as})/2 -i \Gamma_0 (\rho_{aa}-f_{\rm eq})~,
\nn \\
i(\partial_t -Hp\partial_p)  \rho_{ss}
&=& -F_0(\rho_{sa}-\rho_{as})/2~,
\nn \\
i(\partial_t -Hp\partial_p) \rho_{as} &=&
W_0\rho_{as} +F_0(\rho_{ss}-\rho_{aa})/2-
i\Gamma_1 \rho_{as} ~,
\nn\\
i(\partial_t -Hp\partial_p) \rho_{sa} &=& -W_0\rho_{sa} -
F_0(\rho_{ss}-\rho_{aa})/2- i\Gamma_1 \rho_{sa}  ~,
\label{dotrhoall}
\ea
where $a$ and $s$ mean ``active'' and ``sterile'' respectively, $F_0=\dm
\sin 2\theta / 2E$, $W_0= \dm\cos 2\theta /2E + V_{\rm eff}^a$,
$H=\sqrt{8\pi \rho_{\rm tot}/3m_{\rm pl}^2}$ is the Hubble parameter,
\hbox{$m_{\rm pl}=1.22\times 10^{22}$} MeV is the Planck
mass. Furthermore, since the mass of the neutrinos is negligible in
comparison with their momentum, the neutrino momentum and energy
equal, $p=E$. Finally, $f_{\rm eq}$ is the equilibrium Fermi
distribution function, given by
\be
f_{\rm eq}= \frac{1}{ e^{(E-\mu)/T} +1}~,
\label{feq}
\ee
where $T$ is the photon temperature and $\mu$ is the chemical
potential for the active neutrinos.

The effective potential for (anti)-neutrinos is \cite{notzold88}
\be
V_{\rm eff}^a =
\mp C_1 \eta G_FT^3 + C_2^a \frac{G^2_F T^4 E}{\alpha} ~,
\label{nref}
\ee
where $E$ is the neutrino energy, $T$ is the temperature of the
plasma, $G_F=1.166\times 10^{-5}$ GeV$^{-2}$ is the Fermi coupling
constant, $\alpha=1/137$ is the fine structure constant, $C_1=0.345$,
and $C_2^{\rm e}=0.61$ for $\nu_{\rm e}$-$\nu_{\rm s}$ mixing and
$C_2^{\mu, \tau}=0.17$ for $\nu_{\mu,\tau}$-$\nu_{\rm s}$ mixing. The
signs ``$\mp$'' refer to neutrinos and anti-neutrinos, respectively.
The individual contributions to the effective asymmetry $\eta$ from
different particle species are
\ba
\eta &=& 2\eta_{\nue} +\eta_{\num} + \eta_{\nut} +\eta_{\rm
e}-\eta_{\rm n}/2
  ~~~~\mbox{for}~ \nue,
\nn \\
\eta &=& 2\eta_{\num} +\eta_{\nue} + \eta_{\nut} - \eta_{\rm n}/2
  ~~~~\mbox{for}~ \num,
\label{etanumu}\label{etanue}
\ea
and $\eta$ for $\nut$ is obtained from Eq.~(\ref{etanumu}) by the
interchange $\mu \lrar \tau$.

Equations (\ref{dotrhoall}) account exactly for the
first-order terms described by the refractive index, while the second
order terms describing the coherence breaking are approximately modelled by
the damping coefficients $\Gamma_j$.  The latter are equal
to~\cite{bell99}:
\be
\Gamma_0 = 2\Gamma_1  = g_a(p) \frac{180 \zeta(3)}{7 \pi ^4}
\, G_F^2 T^4 p \l[1 \mp z_a \eta_{\nu_{\rm e}} + O(\eta_{\nu_{\rm e}}^2)\r] ~,
\label{gammaj1}
\ee
where the expression in brackets stems from the chemical potential, and
again ``$\mp$'' denotes $\nu$ and $\bar\nu$, respectively. The
coefficients of the expansion in powers of the asymmetry are small,
$z_{\nu_{\rm e}}=0.1$ and $z_{\nu_{\mu, \tau}}=0.04$ \cite{bell99}, and in
the following we will neglect these terms.
The coefficients $g_a(p)$ are in general
momentum-dependent, but in order to simplify
Eqs.~(\ref{dotrhoall}) we can take their
momentum averaged values, which are $g_{\nu_{\rm e}}=3.56$ and $
g_{\nu_{\mu, \tau}}=2.5$
\cite{dolgov99}.

The anti-neutrino density matrix satisfies a set of equations
similar to Eqs.~(\ref{dotrhoall}), but with opposite sign
in the antisymmetric term of $V_{\rm eff}^a$ and with a slight difference
in the damping factors that is proportional to the lepton asymmetry.

For convenience, we will introduce new variables, following the
notation of Dolgov {\it et al.} \cite{dolgov99} to a certain degree,
and we will restrict ourselves to treating $\nu_{\rm e}$ only. First
we note that for $1 \MeV < T< 100 \MeV$ the universe is radiation
dominated so that the scale factor evolves as $a(t) \propto T^{-1}$. This
implies $H = -{\dot T}/T$. Therefore, if we use the comoving frame
given by the variables $x=1\MeV/T$ and $y=p/T$, the LHS of
Eqs.~(\ref{dotrhoall}) simplify to $i x H
\partial_x \rho_{ij}$. Using $\rho_{\rm tot}= \frac{\pi^2}{30}g_\ast
T^4$ with the constant $g_\ast=10.75$, we get $H\simeq 5.5 T^2/m_{\rm
pl}$.

Furthermore, we substitute
\ba
\rho_{aa} &=& f_{\rm eq}(y) [1+a(x,y)]~,\nn \\
\rho_{ss} &=& f_{\rm eq}(y) [1+s(x,y)]~, \nn \\
\rho_{as} = \rho_{sa}^* &=& f_{\rm eq}(y)[h(x,y)+i l(x,y)]~.
\label{hil}
\ea

Next, we introduce a new time variable $q$ which is
defined so that for $\eta=0$ the resonance condition, given by
$W_0=0$, is fulfilled when $q=y$. We find
\be q= \sqrt{\frac{|\dm|\cos 2\theta}{2 C_2^{\rm e} G_F^2 \alpha^{-1}
\MeV^6}}~~ x^3 \approx 6.6\times 10^3 \sqrt{|\dmeV|\cos 2\theta}~x^3~, \ee
where $\dmeV=\dm/\eV^2$. For the evolution of a single momentum mode,
we will use the variable $\tau=q/y$.

With all these modifications, the Eqs.~(\ref{dotrhoall})
have the form
\ba
s' &=& F l~, \label{firstasyms}\nonumber \\
a' &=& - F l - 2 \gamma a ~,\nonumber\\
h' &=& W l - \gamma h  ~,\nonumber\\
l' &=&\frac{F}{2}(a - s) - W h -\gamma l~,
\label{firstasyml}
\ea
where the prime means differentiation with respect to $\tau$. The
coefficients are derived using $W=W_0 \l(Hx \frac{\partial
\tau}{\partial x}\r)^{-1}$, $\gamma=\Gamma_1\l(Hx \frac{\partial
\tau}{\partial x}\r)^{-1}$, and so forth. We also split $W$ into its
symmetric and anti-symmetric parts, so $W = U \mp V \eta$. Then
\be
F= - Q \tan 2\theta \approx -Q \sin 2\theta,\quad \gamma
=\delta/\tau^2,\quad U=Q(1/\tau^{2} -1),\quad V = V_0 y^{-1/3} \tau^{-4/3}
~.
\label{newf}
\ee
The dimensionless constants are given by
\ba
Q &=& \frac{\dm \cos 2\theta}{2 y T}\l(Hx \frac{\partial
\tau}{\partial x}\r)^{-1}
\approx 5.6 \times 10^4\, \sqrt{|\delta m^2| \cos 2\theta}~,
\label{Q}\\
\delta &=& Q ~\frac{g_{\nu_{\rm e}} 90 \zeta(3)}{7 \pi^4
C_2^{\rm e}/\alpha} \approx Q/148~,\\
V_0 &=& \frac{m_{\rm pl}C_1G_F\MeV}{5.5\times3}\l(\frac{|\dm|\cos 2\theta}{2 C_2^{\rm e} G_F^2 \alpha^{-1}
\MeV^6}\r)^{1/6}\approx5.6\times 10^{10} (|\dmeV|\cos 2\theta)^{1/6}~.
\ea
Finally, the effective asymmetry has the form:
\be
\eta = \frac{1}{2\zeta(3)}\int dy ~y^2 f_{\rm eq} ~(a-\bar a) - \eta_0 ~,
\label{etacorr}
\ee
where $\eta_0=\eta_{\rm n}/2-\eta_{\rm
e}-\eta_{\nu_\mu}-\eta_{\nu_\tau}$.

Note that merely $V$ depends directly on the momentum $y$.  Also note
that the system of Eqs.~(\ref{firstasyml}) is completely equivalent to
Eqs.~(\ref{dotrhoall}), i.e.~we have not used any
approximations yet.

\section{Evolution of the lepton asymmetry}

We now briefly discuss the different regimes of the evolution of the
lepton asymmetry. Let us first sketch the reason why the lepton
asymmetry can change at all. Due to the mixing of $\nu_{\rm e}$ with
$\nu_{\rm s}$, the number densities of the $\nu_{\rm e}$ and
$\bar\nu_{\rm e}$ can alter. The sterile neutrinos that are thus
created have no effect on $\eta$. Now if $\eta\neq0$ initially, the
$\nu_{\rm e}$ and $\bar
\nu_{\rm e}$ number densities will change differently,
thus changing $\eta$. Thus the neutrino mixing works as a backreaction
on $\eta$.

\begin{figure}
\unitlength1mm
\begin{picture}(121,80)
  \put(20,0){\psfig{file=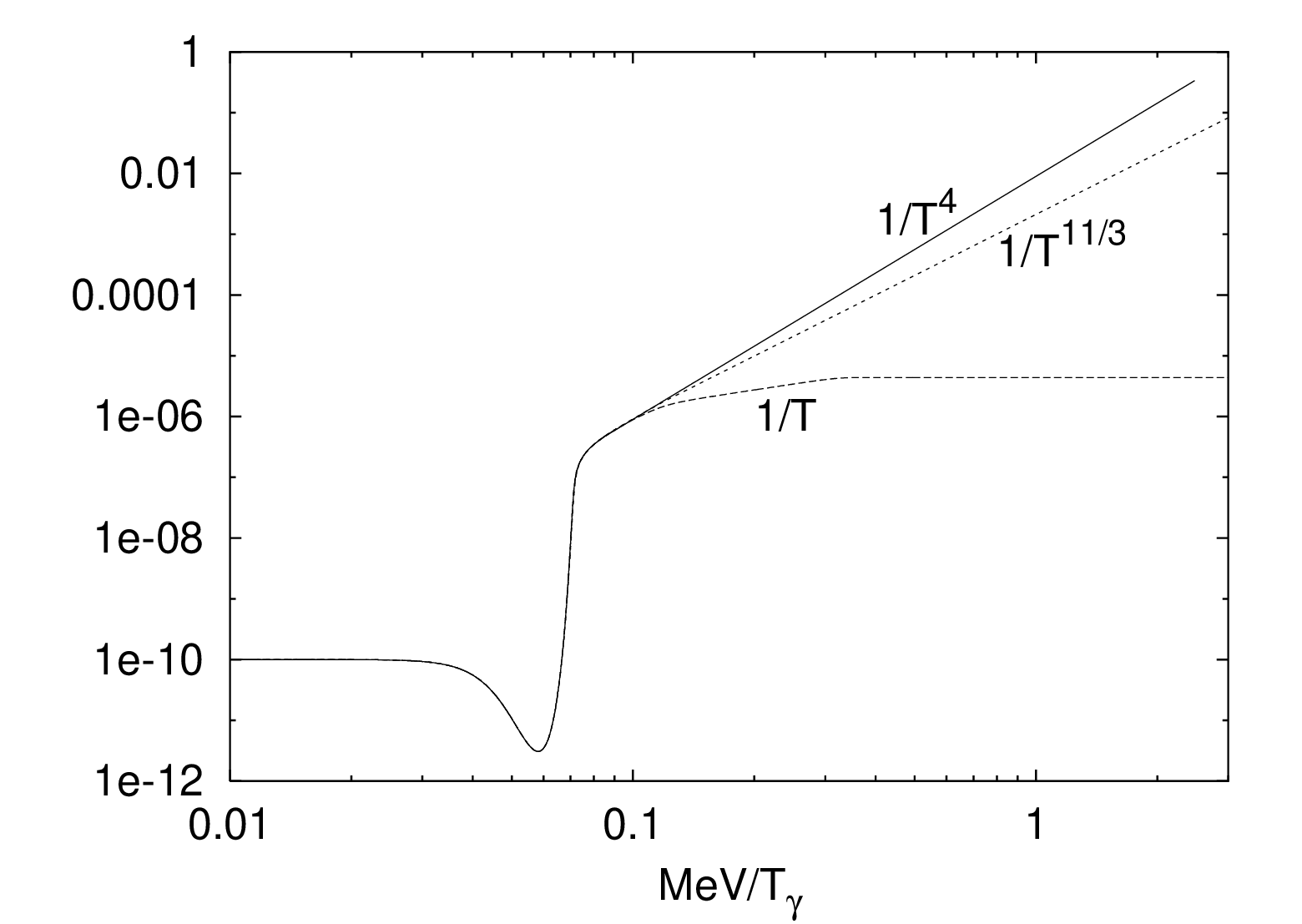,width=10.4cm}} 
  \put(35,75){\footnotesize $\eta$}
  \put(125,8){\footnotesize $x$}
\end{picture}
  \caption{Evolution of the lepton asymmetry. The $T^{-1}$ and $T^{-11/3}$
  lines were calculated numerically with our improved expansion, see
  section (\ref{I123}), with and without the counter-term,
  respectively. For $x>0.3$ we have continued the power-law behavior
  for $T^{-11/3}$, for $T^{-1}$ the power law already freezes out at
  $x\simeq 0.3$. For comparison, we have sketched the $T^{-4}$ power law.}
\label{dPz-evo}
\unitlength1mm
\begin{picture}(121,80)
  \put(20,0){\psfig{file=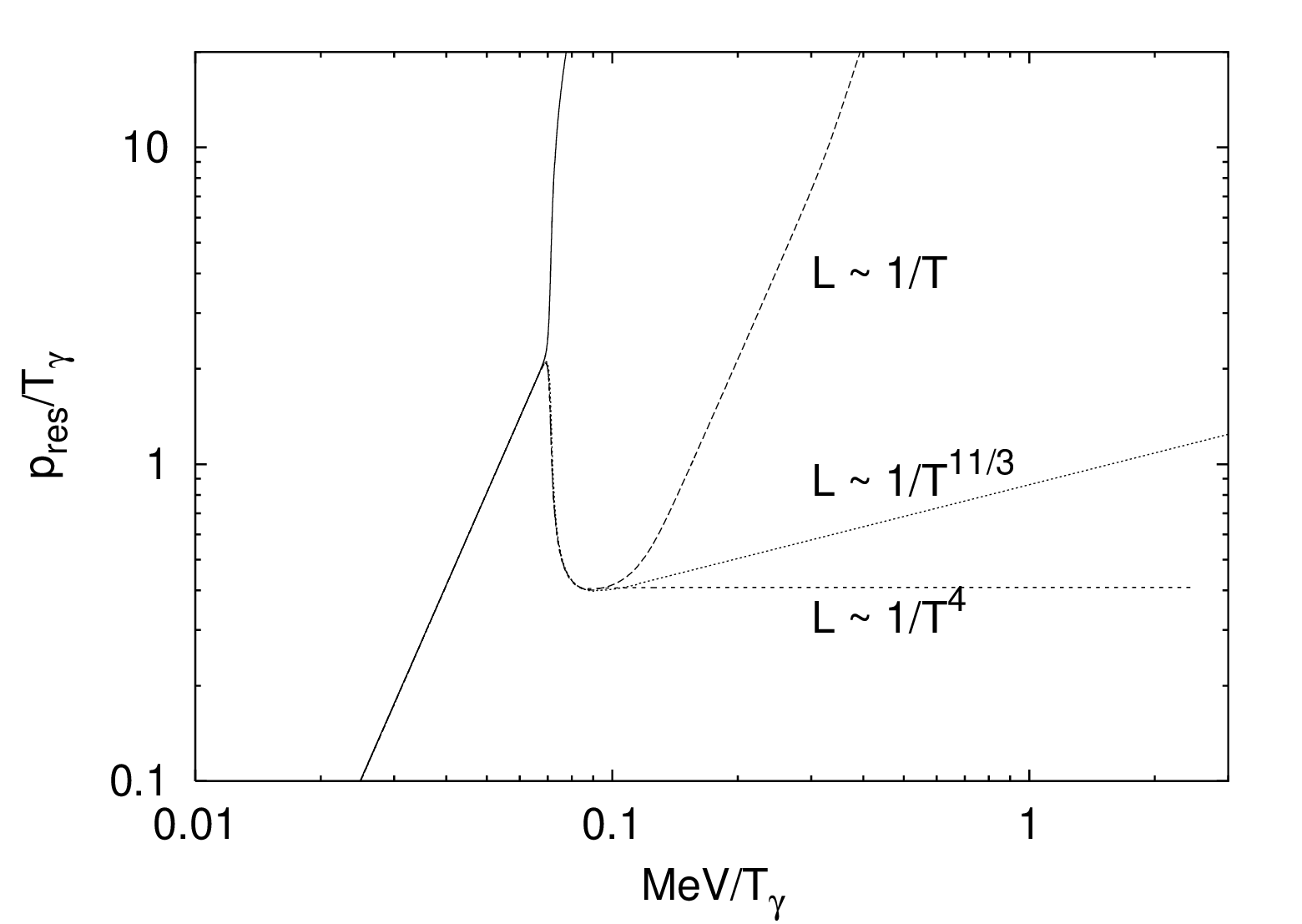,width=10.4cm}} 
  \put(125,8){\footnotesize $x$}
  \put(61,58){$\nu$}
  \put(61,25){$\bar\nu$}
\end{picture}
  \caption{Evolution of the resonance momentum. We treat the same
  cases as in Fig.~\ref{dPz-evo}.}
\label{resmom-evo}
\end{figure}

In Fig.~\ref{dPz-evo} we have plotted an example of the lepton
asymmetry evolution as a function of the inverse background medium
temperature for the case \hbox{$(\delta m^2,\sin 2\theta) =$} \hbox{$(-1 \eV^2,5
\times 10^{-5})$} and an initial effective asymmetry
$\eta_0=-10^{-10}$.  Figure \ref{resmom-evo} shows the evolution of the
resonance momentum $y_{\rm res}$, which is given by
\be
y_{\rm res}= \sqrt{C(q)^2+q^2} \pm C(q)~, 
\label{yres}
\ee
where $C(q)=V_0\eta(q)q^{2/3}/(2Q) $, and $\pm$ is valid for neutrinos
and anti-neutrinos, respectively. For definiteness, we have taken
$\eta$ to be positive.

>From Figs.~\ref{dPz-evo} and \ref{resmom-evo} we can see that there are 
three different stages of lepton asymmetry evolution.
During the first stage the asymmetry is so small that the variable
$C(q)$ can be neglected in Eq.~(\ref{yres}). The neutrinos and the
anti-neutrinos simultaneously pass through the resonance. Furthermore,
any initial asymmetry decreases due to the negative back-reaction.

The second stage is the stage of exponential growth of the lepton
asymmetry, which occurs when the back-reaction becomes positive.  From
the point of view of the resonance condition, the coefficient $C(q)$
becomes important so that the two resonance conditions for $\nu$ and
$\bar\nu$ are driven in opposite directions (see
Fig.~\ref{resmom-evo}). One of them (in our example $\nu$) quickly
passes through all the momentum modes to very high momenta where the
Fermi distribution is negligible, the other one ($\bar\nu$ in our
example) decreases to small momenta $y_{\rm res} < 1$.  Though the
duration of this stage is short in time, the asymmetry grows many
orders of magnitude and reaches values of order of $10^{-6}$.
(For a recent analytical investigation on these two regimes, see
\cite{volkas00-12}.)

However, we should mention that the behavior of the asymmetry in the
exponential regime is still unsettled. Several publications
\cite{chaos} have claimed that the asymmetry starts oscillating in
this regime, thereby changing its sign. We believe that this effect
appears when the momentum distribution is neglected, since then all
neutrinos pass the resonance simultaneously. Still, there might be a
region in the parameter space $(\dm,\sin2\theta)$ where these
oscillations also occur when the momentum distribution is taken into
account \cite{bfoot00}. In the present paper, we will not further
investigate this question and will assume that no such asymmetry
oscillations occur.

In the third stage the lepton asymmetry grows according to a
power law, $\eta\propto T^\alpha$. Unfortunately, in the literature
there exist different results on the exact power $\alpha$. The two
main results are $\alpha=-4$, see e.g.~\cite{foot97}, and $\alpha=-1$,
derived in \cite{dolgov99}.
In the case $\alpha=-4$, the resonance momentum is constant (see
Fig.~\ref{resmom-evo}). Since in this case the $\bar\nu_{\rm s}$ at
the resonance momentum will soon be fully excited, the resonance
momentum of the $\bar\nu$ will in reality slowly increase, thus
passing through the whole spectrum.  Therefore, $\alpha\simgt-4$. In
the end, the $\bar\nu_{\rm s}$ sector will essentially be excited, so
that $\eta={\mathcal O}(0.1)$.
If $\alpha = -1$, the resonance condition passes much faster
through all the momentum modes so that the final asymmetry value
remains small, $\eta \ll 1$. In other words, the $\bar\nu_{\rm s}$ are
only excited to a fractional amount.

The $\eta \propto T^{-1}$ law arrives due to the counter-term in
Eq.~(51) of \cite{dolgov99}. If this counter-term is neglected, we
find a different power law for the asymmetry, $\alpha = -11/3$, a
result which \cite{dolgov99} also found when neglecting their
counter-term. We also show this power law in Fig.~\ref{dPz-evo}. This
power law is quite similar to the one with $\alpha=-4$.

Evidently the final amplitude of the asymmetry strongly depends on
the exact power-law dependence. It is therefore very important to find
the correct answer.

\section{Discussion of the paper by Dolgov et al.}

We now discuss the treatment of the system of differential equations
as done by Dolgov {\it et al.} in \cite{dolgov99}. For convenience, we
will denote references to equations in their paper in the form
(D.51). Furthermore, to avoid confusion, we will strictly use their
notation in this section.

First we would like to pay attention to
some minor misprints, which however do not change the results. First,
the sign in Eq.~(D.20) is wrong. As a consequence, the total sign of
Eq.~(D.51) changes. However, since the initial value of $b_0(0)=+1$
instead of $-1$, the physical content of Eq.~(D.51) remains unchanged.

Next, in Eqs.~(D.47) and (D.48), the signs of the first terms, i.e.~with
the factor $F$, are wrong.

Also, a correct derivation of Eq.~(D.50) yields\footnote{In terms of our
notation, $F$, $\gamma$ and $U$ must be divided by $Q$. Furthermore,
$VZ=D$ corresponds to our $V\eta/Q$. Finally
$\sigma^2=\gamma^2+U^2$ and $\tilde\sigma^2 = \gamma^2 
(1-D^2/\sigma^2)^2 + U^2(1+D^2/\sigma^2)^2$.}
\be
L = 10^{10} { F V Z \gamma U \over \sigma^2 \tilde\sigma^2} b_0\left[
1 + {F^2(\sigma^2 + D^2) \over 4\sigma^2 \tilde\sigma^2}
\left( 1 - 4 \gamma^2 {\sigma^2 + D^2 \over \sigma^2 \tilde\sigma^2} \right)
\right]~,
\label{Lfin}
\ee
where the total sign is the same as in \cite{dolgov99} when taking
into account the changed sign of $Z$. Furthermore, the second term has
changed sign with respect to \cite{dolgov99}, and there
has appeared a new term. Still, the $T^{-1}$ power law holds, since
the third term now acts as the counter-term. To see this, we take the
square brackets in (\ref{Lfin}) at resonance, i.e.~when $D^2=U^2$, and
use the fact that $\gamma^2\ll D^2$, which is certainly true after the
exponential regime. We get $[\dots]= [1+
\frac{F^2}{8\gamma^2}(1-2)]$, which is exactly what we get from an
equivalent treatment of Eq.~(D.50). Thus, the counter-term dominates at
resonance when $\gamma\le|F|/\sqrt{8}$.

As a consequence of Eq.~(20), Eq.~(D.51) becomes
\ba
\frac{1}{Z}\frac{d Z}{dq} &=& - \delta\, B   q^{5/3} \int_0^\infty dt \,
{ t^4 (t^2-1) f_{\rm eq} (tq) b_0(1/t)
\over \sigma^2 \tilde\sigma^2 } \nn\\ &&~~~~~~~~~~~~~~~~~~~~{}\times\left( 1+B_1 { \sigma^2 +D^2 \over
\sigma^2 \tilde\sigma^2 }
\left[1-4 \gamma^2 {\sigma^2 + D^2 \over \sigma^2 \tilde\sigma^2} \right]
\right)~.
\label{dzdq}
\ea
We also checked Eq.~(D.55), and found it to be
\be
\zeta' = \sum_{j=1,2} {\pi B\over 2} b_0\left( 1/t_j\right)
{q^{8/3} (t_j^2 -1) f_{\rm eq} (qt_j) \over 2 \zeta \sqrt{ 4q^{2/3} +\zeta^2}}
\left[ 1 - {B_1 \over 4\delta^2 t_j^4 } \right]~.
\label{zeta'}
\ee
Still, the general physical conclusions do not change, since the
missing factor in the second term \hbox{$(t_j^2-1)^2\simeq1$} anyway. We will show
later on how to derive this type of equation, see Eq.~(\ref{Wexp}ff).

We would also like to comment on the criticism of \cite{dolgov99} given in
\cite{SORRI99}. We agree that neglecting the term $\frac{F}{2}(A-S)$ in
Eq.~(D.28) is not allowed when the asymmetry $Z$ is essentially damped
to zero. However, the effect of this term is merely to slow down the
decrease of $Z$. In fact, it is the asymmetry in chemical potential
that prevents $Z$ from becoming arbitrarily small, see section
\ref{I123}. In any case, during the exponential regime, any information on
the amplitude of $Z$ before the exponential regime is lost. Thus, one
can safely neglect the term in the formalism of \cite{dolgov99}
when one is interested in the final value of the asymmetry.

As a second point, \cite{SORRI99} criticized that Dolgov {\it et al.}
neglected the terms $H'/Q$ and $L'/Q$ in Eqs.~(D.27,D.28). To
check the validity of this approximation, we calculated these two
terms numerically in the formalism used by \cite{dolgov99}, and
compared the results with the other terms in Eqs.~(D.27,D.28). We found
that $H'/Q$ and $L'/Q$ were indeed negligible at all times. The
differing results in \cite{SORRI99} appear due to the neglect of the
momentum distribution. We are of the opinion that one cannot neglect
the momentum distribution, since this would imply that all neutrinos
were at resonance at the same time, which is simply not the case.

Third, the author of \cite{SORRI99} suggested that the $T^{-1}$
power law obtained in \cite{dolgov99} stemmed from neglecting the
influence of the lepton asymmetry on the fermion distribution function
$f_{\rm eq}$. By expanding $f_{\rm eq}$ in $Z\approx \zeta(3)/\pi^2\times 10^{10}
\eta_{\nu_{\rm e}}$, we get
\be f_{\rm eq}(Z) \approx f_{\rm eq}(0)\l( 1 \pm
\frac{Z}{\frac{\zeta(3)}{\pi^2} \times 10^{10}}~ \frac{12 \zeta(3)}{\pi^2}\frac{1}{1+e^{-y}}\r)~,\ee
where $\pm$ denotes neutrinos and anti-neutrinos, respectively. The
correction solely leads to an additional term in $A'/Q$ which is given by
\be -2\gamma Z ~\frac{12}{1+e^{-y}}~. \ee
We compare this with the first term in $A'/Q$, i.e.~$-F L$, using the
leading order in Eq.~(\ref{Lfin}):
\be -2\gamma Z~ \frac{10^{10}F^2 V U}{\sigma^2\tilde\sigma^2}~.\ee
One can clearly see that the additional term can safely be neglected
near the resonance in the power-law regime.

Finally, we discuss the problems we found in the analysis of
\cite{dolgov99}. The first occurred when the authors diagonalized their
set of equations for the symmetric functions, Eq.~(D.33). When
diagonalizing and expanding it in $F$, they get as the second
eigenvalue
\be \mu_2 \approx -2\gamma + \frac{F^2 \gamma}{2[\sigma^2-(VZ)^2]}~, \ee
where $\sigma^2=\gamma^2+U^2$. Here, the denominator of the second
term diverges close to the resonance in the case that $V Z\ge \gamma$.
We found that this condition is already fulfilled before the end of
the exponential regime. Furthermore, we found that this divergence
continued to cause problems in the subsequent analysis.

The second problem occurs when $F={\mathcal O}(\gamma)$. Then the
expansion in $F$ is no longer valid. Unfortunately, the counter-term in
Eq.~(D.50), i.e.~our Eq.~(\ref{Lfin}), becomes of
order 1 only when $F\approx \gamma$, so that one cannot say whether
the counter-term is physical or simply an artifact of the expansion.

Finally, the validity of Eq.~(D.49) breaks down when
$Q\gamma\ll1$. Thus, assuming that the counter-term is of order 1,
(D.49) is invalid for $QF\ll1$, which corresponds to
$\sin2\theta\sqrt{|\dm|\cos2\theta}\ll1.8 \times 10^{-5}$.

\section{Improved expansion}

In this section we develop a similar approach to that of
\cite{dolgov99}, using our own notation
unless quoted explicitly. However, we want to circumvent the divergent
behavior that occurs in their analysis. To this end, we treat the
neutrinos and anti-neutrinos separately.

The system of equations (\ref{firstasyml}) can be rewritten in
matrix form as ${\cal V}' = {\cal M} {\cal V}$:
\ba
\left( \begin{array}{c} s' \\ a'\\ h'\\ l' \end{array} \right)
= \left( \begin{array}{cccc}
0  & 0 & 0 & F\\
0  & -2 \gamma & 0  & -F\\
0 & 0 & -\gamma &  W\\
-F/2 & F/2 &  - W & - \gamma    \end{array}  \right)
\left( \begin{array}{c} s \\ a \\ h \\ l  \end{array} \right)~.
\label{symmatrix}
\ea
We can apply a non-degenerate transformation 
\be
{\cal V}={\cal X} {\cal B}~.
\label{change_var}
\ee
For the new vector ${\cal B}$ we get the equation
\be
{\cal B}' = \left( {\cal X}^{-1}{\cal M}{\cal X} +  {\cal X}^{-1}  {\cal X}'\right) {\cal B}~.
\label{gauge}
\ee
Now, if ${\cal X}$ is a matrix, constructed from the eigenvectors of
${\cal M}$, then the first term on the RHS of Eq.~(\ref{gauge}) will
be diagonal.  In the case where ${\cal M}$ is a constant matrix, the
eigenvectors will also be constants and the second term on the RHS of
Eq.~(\ref{gauge}) will be zero. Then the equations for each component
of ${\cal B}$ will be independent of each other and could easily be
solved. Then the general solution of the matrix equation will be a
linear combination of the eigenvectors of ${\cal M}$ with coefficients
${\cal B}$.

In the general case, when ${\cal M}$ is not constant, there is no
straightforward way to solve the matrix equation.  But in our case in
some regimes we can hope that the second term on the RHS of
Eq.~(\ref{gauge}) will be a small correction in comparison to the main
term, and then we can solve Eq.~(\ref{gauge}) perturbatively.

To begin, we write down the exact eigenvalues of ${\cal M}$:
\ba
\mu_1 &=& -\gamma -\frac{1}{\sqrt{2}} \sqrt{-F^2 - W^2 + \gamma^2 +
              \sqrt{-4 F^2\gamma^2 + (F^2 +W^2 +\gamma^2)^2}}~,
\nonumber \\
\mu_2 &=& -\gamma +\frac{1}{\sqrt{2}} \sqrt{-F^2 - W^2 + \gamma^2 +
              \sqrt{-4 F^2\gamma^2 + (F^2 +W^2 +\gamma^2)^2}}~,
\nonumber \\
\mu_3 &=& -\gamma -\frac{1}{\sqrt{2}} \sqrt{-F^2 - W^2 + \gamma^2 -
              \sqrt{-4 F^2\gamma^2 + (F^2 +W^2 +\gamma^2)^2}}~,
\nonumber \\
\mu_4 &=& -\gamma +\frac{1}{\sqrt{2}} \sqrt{-F^2 - W^2 + \gamma^2 -
              \sqrt{-4 F^2\gamma^2 + (F^2 +W^2 +\gamma^2)^2}}~.
\label{eigenvalues}   
\ea
Note that at resonance $W=0$, while out of resonance $W \gg F,
\gamma$. We will consider three different regimes, $F \ll \gamma$, $F
\sim \gamma$ and $F \gg \gamma$. We will concentrate on the first
regime and will afterwards stress the problems occurring in the two
other regimes.

\subsection{Expansion for \boldmath{$F \ll \gamma$}}

In this case we expand Eqs.~(\ref{eigenvalues}) in $F$. We have done
so to second order:
\ba
\mu_1 &=& - 2 \gamma  + \frac{F^2 \gamma}{2 \rho^2}~,
\nonumber \\
\mu_2 &=& - \frac{F^2 \gamma}{2 \rho^2}~,
\nonumber \\
\mu_{3,4} &=& - \gamma \mp i W \left(1 + \frac{F^2}{2\rho^2}\right)~,
\label{eigen_F}
\ea
where we have used $\rho^2=\gamma^2+W^2$.

As a next step, we can construct the eigenvectors, and thus the matrix
${\cal X}$, from them.  Then we re-express the original wave function
vector $(s,a,h,l)$ through the new functions $b_j(\tau)$ according to
(\ref{change_var}), again keeping only terms up to order $F^2$:
\ba
s &=& -\l(1-\frac{F^2(-3W^2+\gamma^2)}{2\rho^4}\r) b_1 +b_2 {F^2\over 4\rho^2} \nonumber \\
&&- {F \over \rho^2}
\left[\gamma \left( b_3 \cos \Omega -b_4 \sin \Omega\right) -
 W \left(b_3 \sin \Omega + b_4\cos \Omega\right)\right] ~,
\label{sb}  \\
a &=& - b_1 {F^2 \over 4 \rho^2} +
\l(1-\frac{F^2(-3W^2+\gamma^2)}{2\rho^4}\r)b_2 \nonumber \\
&&-{F\over \rho^2}\left[\gamma \left( b_3 \cos \Omega -b_4 \sin \Omega\right)
+W \left(b_3 \sin \Omega + b_4\cos \Omega\right)\right] ~,
\label{ab}  \\
h &=& b_1 {FW \over 2 \rho^2} +
b_2 {FW \over 2 \rho^2} +
 \left(b_3 \sin \Omega + b_4\cos \Omega\right)~,
\label{hb} \\
l &=& b_1 {F\gamma \over 2 \rho^2} -
b_2 {F\gamma \over 2 \rho^2} +
 \left(b_3 \cos \Omega -b_4 \sin \Omega\right)~.
\label{lb}
\ea
Here, we have at the same time introduced the fast-oscillating
variable $\Omega$, with $\Omega' = W$. Also, while the functions $b_1$
and $b_2$ are the eigenfunctions of the eigenvalues $\mu_1$ and
$\mu_2$, respectively, the functions $b_3$ and $b_4$ are
superpositions of the eigenfunctions to the remaining eigenvalues: the
actual eigenfunctions are $b_2(\cos\Omega+i
\sin\Omega)+$ \hbox{$b_3(-\sin\Omega+i\cos\Omega)$} and $b_2(\cos\Omega-i
\sin\Omega)+b_3(-\sin\Omega-i\cos\Omega)$ for $\mu_3$ and $\mu_4$,
respectively. We have chosen these definitions to circumvent imaginary
values.

Inserting these definitions into the system of differential equations
(\ref{symmatrix}) we derive
\ba b_0'&=&-\frac{F^2 \gamma}{2\rho^2}b_0~, \\
    b_2'&=&-\gamma b_2 -\ot b_0 \l[\l(\frac{FW}{\rho^2}\r)'\sin\Omega
    + \l(\frac{F\gamma}{\rho^2}\r)'\cos\Omega\r]~, \\
    b_3'&=&-\gamma b_3
    -\ot b_0 \l[\l(\frac{FW}{\rho^2}\r)'\cos\Omega -
    \l(\frac{F\gamma}{\rho^2}\r)'\sin\Omega\r]~,
\ea
where we have neglected all terms of insignificant order, and $b_1$ can
be neglected totally. Furthermore, the initial values are given by
$b_0(0)=+1$ and $b_{1,2,3}(0)=0$. Following the corrected
Eqs.~(D.45--D.51), we get
\be l(\tau) =
\frac{F\gamma}{2\rho^2}b_0\l[1+\frac{F^2}{2\rho^2}\l(1-\frac{2\gamma^2}{\rho^2}\r)\r]~. \label{lour}\ee
Also in this case we find a counter-term which is similar to that
derived by \cite{dolgov99}.

To obtain the evolution of the lepton asymmetry, we also have to
include the anti-neutrinos, i.e.
\be \bar l(\tau) =
\frac{F\bar\gamma}{2\bar\rho^2}\bar b_0
\l[1+\frac{F^2}{2\bar\rho^2}\l(1-\frac{2\bar\gamma^2}{\bar\rho^2}\r)\r]~.
\label{bar_l}
\ee

Then, by using the differential equation for $a$ and
Eq.~(\ref{etacorr}), we get
\be \frac{d\eta}{dq} = \frac{|F|}{2\zeta(3)}\int_0^\infty
dy~ y \l[f_{\rm eq}(y, \mu) l\l(\mbox{$\frac{q}{y}$}\r)-f_{\rm eq}(y,\bar
\mu)\bar
l\l(\mbox{$\frac{q}{y}$}\r)\r]~. \label{detadq} \ee
Note that one power of $y$ cancels when we change from $d/d\tau$ to
$d/dq$.  Furthermore, the $\gamma a$ term in the differential equation
for $a$ disappears due to the conservation of leptonic charge
\be \int dy~y \l[ f_{\rm eq}(y, \mu) \gamma a-f_{\rm eq}(y, \bar
\mu)\bar\gamma\bar a\r]=0~. \label{conslc}\ee

Equation (\ref{detadq}) is valid in the case $F < \gamma$. For $F\ge
\gamma$, the perturbative expansion breaks down. We should also check the
validity of Eq.~(D.49), i.e.~the approximation of the integral. This
approximation is valid when $\gamma\gg 1$, which is a weaker
condition than $F<\gamma$ if  $F>1$, i.e.~$\sin 2 \theta
\sqrt{|\dmeV|\cos 2 \theta} > 1.8\times 10^{-5}$. Thus, in the case of small
mixing angles and small $\dm$, the condition that Eq.~(\ref{detadq})
is valid is given by $\gamma > 1$. From solving Eq.~(\ref{detadq})
numerically, we found that these conditions hold until the power-law
regime is established for $\sin 2 \theta < 5 \times 10 ^{-5}$ when $\delta
m^2 = -1 \eV^2$. If we neglect the counter-term in Eq.~(\ref{detadq}),
we get the same solution as \cite{foot97}, Eqs.~(79-80), up to some
minor corrections.

To compare our result Eqs.~(\ref{lour}--\ref{detadq}) with
\cite{dolgov99}, we neglect the chemical potential $\mu$, use
$b_0\approx\bar b_0$ and expand $(l-\bar l)$ in $V \eta$. We get
\ba (l-\bar l)&\simeq& \frac{2F\gamma
UV\eta}{\sigma^4}b_0\l(1+\frac{F^2(U^2-2\gamma^2)}{\sigma^4}\r. \\&&~~~~~~~~~~~~~~~~~{}+\l.(V\eta)^2
\l\{\frac{2(U^2-\gamma^2)}{\sigma^4}
+\frac{F^2(5U^4-26U^2\gamma^2+9\gamma^4)}{\sigma^8} \r\}\r)~, \nn\ea
where $\sigma^2=U^2+\gamma^2$. We compare it with the expansion of
(\ref{Lfin})
\ba 2\times 10^{-10} L&\simeq& \frac{2F\gamma
UVZ}{\sigma^4}b_0\l(1+\frac{F^2(U^2-3\gamma^2)}{4\sigma^4}\r. \\&&~~~~~~~~~~~~~~{}+\l.(VZ)^2
\l\{\frac{2(U^2-\gamma^2)}{\sigma^4}
+\frac{F^2(5U^4-30U^2\gamma^2+13\gamma^4)}{4\sigma^8}
\r\}\r)~. \label{tdo}\nn\ea
This equation is given in terms of the variables as defined in
\cite{dolgov99}. Still we can compare the two equations, since the
differences to our definitions cancel.  We see that, up to some minor
corrections, the main difference is in a factor of $4$ in the $F^2$
terms. This factor has no fundamental consequences.

\subsection{Discussion of \boldmath{$F\ll\gamma$}}
\label{I123}

Now we discuss the solutions of Eq.~(\ref{detadq}) in the region of
its validity in more detail.  First, let us consider the region where
the asymmetry decreases (see Fig.~\ref{dPz-evo}). In this region, the
corrections to the main terms in Eqs.~(\ref{lour}) and (\ref{bar_l})
are negligible. We expand the integrand of Eq.~(\ref{detadq}) in terms
of $\eta$ and $\mu/T$ and get
\ba
&& f_{\rm eq}(y, \mu) l - f_{\rm eq}(y, -\mu) \bar l \approx
f_{\rm eq}(y) \frac{F \gamma_0 b_+}{2} \frac{4 U V \eta}{\rho^2 \bar \rho^2}
- f_{\rm eq}(y) \frac{F \gamma_0 b_-}{2} \l( \frac{1}{\rho^2} +
\frac{1}{\bar\rho^2}\r)
\nonumber \\
&+& f_{\rm eq}(y) \frac{F \gamma_0 b_+ \eta_{\nu_{\rm e}}}{2}\l( \frac{1}{\rho^2} +
\frac{1}{\bar\rho^2}\r)
\l\{3[1-f_{\rm eq}(y)]-0.2 \r\} + O(\eta_{\nu_{\rm e}}^2, \eta^2,
\eta_{\nu_{\rm e}}\eta)~,
\label{exp_l}
\ea
where we used the notations
\be
b_+ = \frac{b_0 +\bar b_0}{2} ~~~~\mbox{and}~~~~ b_-= b_0 - \bar b_0~.
\ee
On the RHS of Eq.~(\ref{exp_l}), the first term corresponds to the
main term in Eq.~(D.50), the second term takes into account the $A-S$
contribution neglected in Eq.~(D.50), and the last term takes into
account the non-zero chemical potential both for the $\gamma$-terms,
see Eq.~(\ref{gammaj1}), and for $f_{\rm eq}$, which we rewrite through
$\eta_{\nu_{\rm e}} \approx 1.5 \frac{\mu}{T}$. Note that we have used
$\bar\mu=-\mu$, which is valid for large temperatures. Before
substituting Eq.~(\ref{exp_l}) into Eq.~(\ref{detadq}), let us write
the differential equations for $b_\pm$, in which we can neglect the
small difference between $\gamma$ and $\bar\gamma$:
\ba
\frac{db_+}{d\tau} &=& -\frac{F^2 \gamma}{2}\l(
\frac{\gamma^2 + U^2 + (V\eta)^2}{\rho^2 \bar \rho^2}b_+ +
\frac{UV\eta}{\rho^2 \bar \rho^2}b_-
\r)~,
\nonumber \\
\frac{db_-}{d\tau} &=& -\frac{F^2 \gamma}{2}\l(
\frac{\gamma^2 + U^2 + (V\eta)^2}{\rho^2 \bar \rho^2}b_- +
\frac{4 UV\eta}{\rho^2 \bar \rho^2}b_+
\r)~.
\ea
In the case of small asymmetry $\eta$ we approximately get from these
equations $b_+ \approx 1$ and $b_- \propto \sin^2 2\theta\eta$. Using
this fact, we can substitute Eq.~(\ref{exp_l}) into Eq.~(\ref{detadq})
and get the following equation:
\be
\frac{d\eta}{dq} = - C \eta \l[ I_1(q) - \sin^2 2\theta ~I_2(q) \r] +
\eta_{\nu_{\rm e}}
I_3(q)~,
\label{small_eta}
\ee
where the functions $I_i(q)$ are integrals over the momentum which do
not depend on the asymmetries. The term $I_1(q)$ corresponds to the
main term in (D.51), and without $I_2$ and $I_3$ the asymmetry
decreases to infinitely small values.

The second term has opposite sign and is proportional to an additional
power of $\sin^2 2\theta $. For $\sin^2 2\theta \sim 10^{-4}$, this
term becomes of the order of the first one, but it can not cancel it
completely. We would like to mention that, although this term seems to
be related to the expansion in $F$, it is actually of a different
nature, and should therefore not be neglected here. This term
corresponds to the $A-S$ contribution in terms of \cite{dolgov99} and
it is proportional to $b_-$ in our notations.

But the actual reason that the lepton asymmetry does not decrease to
very small values is the non-zero chemical potential, which
corresponds to $I_3$. When the asymmetry, which evolves according to
Eq.~(\ref{small_eta}), becomes small enough, $\eta = 2 \eta_{\nu_{\rm e}}
-\eta_0 \ll \eta_{\nu_{\rm e}}\approx -\eta_0/2$, the last term in
Eq.~(\ref{small_eta}) starts to compensate the first two terms, and
prevents the asymmetry to get tiny values.

We will not consider the regime of exponential growth of the asymmetry
(see Fig.~\ref{dPz-evo}) separately, let us just note again that
during this stage the resonance condition for anti-particles
(particles) go to very large momenta, while the resonance condition
for particles (anti-particles) go to small momenta, $y<1$, for
$\eta>0$ ($\eta<0$), see Fig.~\ref{resmom-evo}.  Therefore, in the
power-law regime, one may neglect either neutrinos or
anti-neutrinos in (\ref{detadq}), since the contribution of one of
them is negligible due to the exponential damping of the Fermi
distribution function.

Finally, we discuss the power-law regime. In this regime, we can
neglect either neutrinos or anti-neutrinos. For definiteness, we
consider the case $\eta<0$, where the resonance momentum of the
anti-neutrinos is very large. Thus we can neglect $\bar l$ in
Eq.~(\ref{detadq}). Also, we will assume that the asymmetry is still
$\eta\ll1$, so that we can also neglect the chemical potential. This
is in contrast to the case described by Eq.~(\ref{small_eta}), where
the main terms of the $\nu$ and $\bar\nu$ almost canceled each other,
so that the even the small chemical potential would become important. Here,
however, the $\bar\nu$ have no influence at all, so that no
cancellation of the main terms appears.
Furthermore, we assume $b_0\simeq1$, which is valid for small
$F/\gamma$. To be able to integrate analytically, we expand $W$ around
$y_{\rm res}$, given in Eq.~(\ref{yres}):
\be W(y)\approx c_W (y-y_{\rm res})+\dots, ~~~~\mbox{with}~~~~ c_W =-\frac{2y_{\rm
res}Q}{q^2}+V\eta~. \label{Wexp}\ee
Then the leading order in (\ref{detadq}) gives
\ba \frac{d\eta}{dq} &=& \frac{-F^2 \delta}{4\zeta(3)q^2}
\int\limits_0^\infty dy f_{\rm eq}(y)\frac{y^3}{\gamma^2+W^2} \nn\\
&\approx&{}
\frac{-F^2 \delta}{4\zeta(3)q^2}f_{\rm eq}(y_{\rm res})\frac{y_{\rm
res}^3}{\gamma^2 (y_{\rm res})}
\int\limits_{-\infty}^{\infty} dy \frac{1}{1+\frac{c_W^2}{\gamma_{\rm
res}^2}
(y-y_{\rm res})^2}\nn \\ &=&{} \frac{-F^2 \delta}{4\zeta(3)q^2}f_{\rm eq}(y_{\rm res})\frac{y_{\rm
res}^3}{\gamma^2 (y_{\rm res})}
\frac{\gamma(y_{\rm res})}{c_W}\int\limits_{-\infty}^\infty dz \frac{1}{1+z^2}~,\label{lot}\ea
where we have substituted $z=c_W(y-y_{\rm res})/\gamma_{\rm res}$. The
final integral gives $\pi$. Analogously, we find the next order
corrections, so that we get
\be \frac{d\eta}{dq}=-\frac{F^2\pi}{4\zeta(3)}f_{\rm eq}(y_{\rm res})\frac{y_{\rm res}}{c_W}
\l( 1 - \frac{F^2}{8\gamma^2(y_{\rm res})}\r)~, \label{detadqint}\ee
which is our differential equation for the asymmetry $\eta$ in the
power-law regime. We can compare this to Eq.~(\ref{zeta'}), which is
the corresponding equation for the approach by Dolgov {\it et al.}. The
interesting point is that the counter-term, compared with the leading
term, is the same in both approaches up to a factor of
2 (the counter-term in Eq.~(\ref{zeta'}) corresponds to
$F^2/(16\gamma^2(y_{\rm res}))$ in our notation).

>From Eq.~(\ref{detadqint}) it is easy to derive the power law for
$\eta$. Let us first neglect the counter-term.
If we further assume that $C(q)\gg q$ in (\ref{yres}), we get $y_{\rm
res}\approx Q/(V_0\eta q^{-4/3})$ and can neglect the first term in
$c_W$. Then
\be \frac{d\eta}{dq}\approx \frac{-F^2\pi}{4\zeta(3)}f_{\rm eq}(y_{\rm
res}) \frac{Q}{V_0^2}\eta^{-2}q^{8/3}\equiv C_\eta \eta^{-2}q^{8/3}~.
\label{defCeta}\ee
If we multiply by $dq$ and integrate, we get
\be \eta = \l(\frac{9}{11}C_\eta\r)^{1/3}q^{11/9}~.\label{derive119} \ee
Here, we have neglected the boundary conditions, which become
negligible for large $q$ in any case. We see that this corresponds to
the $\eta\propto T^{-11/3}$ power law found in \cite{dolgov99} when
neglecting the counter-term. We should however mention that even for
all the simplifications made above, this power law is only
approximate, since $C_\eta$ contains the Fermi distribution $f_{\rm
eq}(y_{\rm res})$. Since $y_{\rm res}\propto q^{1/12}$ for the
derived power law, $f_{\rm eq}(y_{\rm res})$ will slowly decrease,
thus changing Eq.~(\ref{derive119}).

If we take the counter-term into account, the
condition for the power law is that the brackets in Eq.~(\ref{detadqint})
equal 0. Using again the approximation $y_{\rm res}\approx
Q/(V_0\eta q^{-4/3})$, we can derive
\be \eta = \frac{Q}{V_0}\l(\frac{8\delta^2}{F^2}\r)^{1/4}q^{-1/4}~, \ee
which is exactly the $\eta\propto T^{-1}$ power law found in
\cite{dolgov99}.

\begin{figure}
\unitlength1mm
\begin{picture}(121,80)
  \put(20,0){\psfig{file=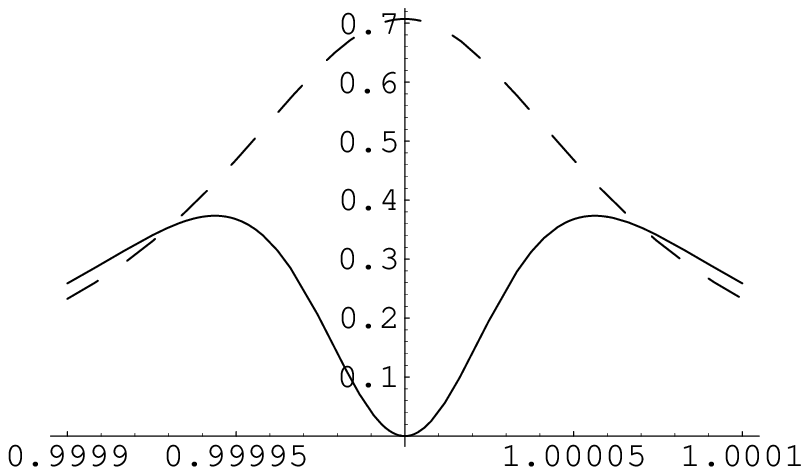,width=10.4cm}} 
  \put(68,68){$l$}
  \put(130,5){$\frac{y}{y_{\rm res}}$}
\end{picture}
  \caption{\label{count-plot1} $l(y)$ for the case $F=\sqrt{2}\gamma$
  at resonance and $\eta\propto T^{-1}$. The abscissa shows the momentum
  relative to the resonance momentum, i.e.~$y/y_{\rm res}$. The dashed
  line shows the first term in Eq.~(\ref{lour}), while the solid line
  shows the full equation.}
\end{figure}

Let us still have a look on the evolution of $l$ for the $\eta\propto
T^{-1}$ power law. In Fig.~\ref{count-plot1} we have plotted $l$ as
given in Eq.~(\ref{lour}), once with and once without the counter-term,
for the case where $F=\sqrt{2}\gamma$ at resonance. Then the
counter-term merely compensates the leading order term near the
resonance, but the integration over the momentum still yields a
similarly large value when taking the counter-term into account.  Thus,
it is safe to say that the counter-term is of minor importance for
$F\le\sqrt{2}\gamma$. This corresponds to the result in
Eq.~(\ref{detadqint}).

To summarize, this ansatz gives similar results as the approach of
\cite{dolgov99}, so that the divergence in their approach seems not to
be critical. However, our approach has shown two further advantages:
First, we need not make the adiabatic approach which they used
for $L$ and $H$. Second, our result yields the correct behavior for
very small $\eta$ before the exponential regime, since we included the
term $(F/2) (a-s)$ and the chemical potential.

\subsection{\boldmath{$F \gg \gamma$}}

This region can occur in two cases: either when $F$, and thus the
mixing angle $\sin 2 \theta$, is large, or when $\gamma$ is very
small, which is the case for $\tau\gg1$. Since we only discuss small
mixing angles in our analysis, the first case is not
important. However, the second case appears, since for large
asymmetries, the (anti-)neutrinos at resonance have $\tau=q/y$, where
$q>1$ is large and $y\sim 0.1$ is small.

In this limit the eigenvalues Eqs.~(\ref{eigenvalues}) should be
expanded in $\gamma$. However, in this regime, the approximation
in Eq.~(D.49) does not hold, since $\gamma$ is no
longer very large. Therefore, we did not investigate further in this
case.

\subsection{ \boldmath{$F \sim \gamma$} }

In this case let us rewrite $\gamma = F (1 + \chi)$ and suppose
that $\chi \ll 1$. 
Then we can expand the eigenvalues (\ref{eigenvalues})
in $\chi$. To the main order, which does not contain $\chi$, we get:   
\be
\mu_i = -F \pm \frac{1}{\sqrt{2}} \sqrt{-W^2 \pm W \sqrt{W^2+ 4 F^2}}+O(\chi)~.
\label{eigen_gF}
\ee
The important feature of Eq.~(\ref{eigen_gF}) is the fact that at
resonance, $W=0$, all 4 eigenvalues coincide. This means that the
resonance matrix ${\cal M}$ changes its rank, and thus diagonalization
becomes impossible.

We conclude that the approach of Dolgov {\it et al.} is useful and
correct as long as $F<\gamma$, but that it seems very difficult to
develop this kind of approach for $F\simgt\gamma$.


\section{Solution for \boldmath{$|a-s-1|\ll1$}}

\subsection{Solution for one momentum mode}

In this section, we present a different analytic solution of the
system of differential equations (\ref{firstasyml}) in the power-law
regime for a given momentum $y$. We introduce two
simplifications. First, to circumvent the non-linearity, we give the
lepton asymmetry an explicit form,
\be \eta(T)=\eta_i T^\alpha~, \ee
where $\alpha$ represents the power-law behavior. Second, we treat
the change of the variables $a$ and $s$, expressed by the function
$\fas\equiv(a-s-1)$, as a perturbation in the differential equation of
$l$. In terms of the physical content, $f_{(a-s)}\ll1$ means that
there is only very little conversion from active to sterile
neutrinos. The advantage of this approximation is simple: provided
that the $T^{-1}$ power law holds, the condition $f_{(a-s)}\ll1$ holds
for all times\footnote{This condition corresponds to $P_z-1\ll1$ in
the formalism used by \cite{foot97} and others.}.

To start with, we write down the differential equations for $h$ and
$l$:
\ba l'&=&\frac{F}{2}(1+\fas)-W h-\gamma l~, \label{leqorig} \\
    h'&=&                    W l-\gamma h~. \label{heqorig} \ea
Subsequently, we will assume that $\fas$ is an explicit function. Then
the above equations form a complete set of differential equations for
$h$ and $l$.

By solving (\ref{leqorig}) for $h(\tau)$, taking its derivative $h'(\tau)$, and
inserting both into (\ref{heqorig}), we find the differential equation
of second order for $l(\tau)$:
\ba W l''+\l(2W\gamma-W'\r)l' + \l(W^3+\gamma^2W+W\gamma'-\gamma W'\r)l && \nn\\
+ \frac{F}{2}\l[(1+\fas)\l(W' - \gamma W\r)-W \fas'\r] &=&0~. \label{delex}\ea
Analogously, we find
\ba W h''+\l(2W\gamma-W'\r)h' + \l(W^3+\gamma^2W+W\gamma'-\gamma W'\r)h && \nn\\
- \frac{F}{2}(1+\fas) W^2 &=&0~, \ea
where the only difference to (\ref{delex}) is given in the
inhomogeneous $F$-term.

We can now solve the equation for $l$ analytically. To this end, we
first write down the solution for the homogeneous differential
equation, i.e.~$F=0$:
\ba l_{\rm hom}(\tau) &=& \l(k_c^l \cos(\int\limits_{\tau_i}^\tau W_2d\tau_2)
                  + k_s^l \sin(\int\limits_{\tau_i}^\tau
W_2d\tau_2)\r)\nn\\ &&{}\times \exp\l(-\int\limits_{\tau_i}^\tau
\gamma_2d\tau_2\r)~.\label{lhom} \ea
Here, the coefficients $k_c^l$ and $k_s^l$ are given by the boundary
conditions at time $\tau_i$, and the subscript 2 means that the
parameters are evaluated at $\tau_2$. The solution for the
inhomogeneous differential equation can then easily be found, e.g.~by
using \cite{kuk},
\ba l_{\rm inhom}(\tau)&=& \frac{F}{2}\int\limits_{\tau_i}^{\tau}
\frac{(1+\fas)_1\l(W_1' - \gamma_1 W_1\r)-W_1 (\fas')_1}{W_1^2} \nn\\
&&{}~~~~~~~\times
\sin\l(-\int\limits_{\tau_1}^{\tau}W_2d\tau_2\r)
\exp\l(-\int\limits_{\tau_1}^{\tau}\gamma_2d\tau_2\r)d\tau_1~. \label{linhom}\ea
The complete solution is then $l_{\rm analytic}=l_{\rm hom}+l_{\rm
inhom}$.

Analogously, we get $h=h_{\rm hom}+h_{\rm inhom}$, where $h_{\rm hom}$
is identical to the homogeneous solution for $l$, (\ref{lhom}), now
with coefficients $k_c^h$ and $k_s^h$. The inhomogeneous solution is
\be h_{\rm inhom} = \frac{F}{2}\int\limits_{\tau_i}^{\tau}
(1+\fas)_1\sin\l(-\int\limits_{\tau_1}^{\tau}W_2d\tau_2\r)
\exp\l(-\int\limits_{\tau_1}^{\tau}\gamma_2d\tau_2\r)d\tau_1~. \label{hinhom}\ee

Now we turn to the variables $a$ and $s$. It is straightforward to
solve the differential equations describing them, see (\ref{firstasyml}),
provided that $l$ is an explicit function:
\ba a(\tau)&=&a_i \exp\l(-\int\limits_{\tau_i}^{\tau}\gamma_2d\tau_2\r)
- F \int\limits_{\tau_i}^{\tau}\lexpl(\tau_1)
 \exp\l(-\int\limits_{\tau_1}^{\tau}\gamma_2d\tau_2\r)d\tau_1 \label{aanal}\\
s(\tau)&=&s_i + F \int\limits_{\tau_i}^{\tau}\lexpl(\tau_1)
 d\tau_1~. \label{sanal}\ea

We still need to settle the boundary conditions: From \cite{dolgov99}
we know that for $\tau_i=0$, $h=l=a=0$ and $s=-1$. It follows that
$k_c^l=k_s^l=k_c^h=k_s^h= a_i=0$ and $s_i=-1$ as initial
conditions.

The next step is to connect the two systems of differential equations,
$(h,l)$ and $(a,s)$, using perturbation theory. The leading order,
labeled with the subscript $0$, is defined by
$\fas\equiv\lexpl\equiv0$. We find $a_0(\tau)\equiv s_0(\tau)\equiv
0$, while $l_0$ and $h_0$ are given by (\ref{linhom}) and
(\ref{hinhom}), respectively, using $\fas\equiv\fas'\equiv0$.

The first-order corrections, $a_1$ and $s_1$, are found by inserting
$\lexpl=l_0$ into the analytic solutions, (\ref{aanal}) and
(\ref{sanal}), respectively, and for $h$ and $l$ by inserting
$\fas=a_1-s_1-1$ into (\ref{hinhom}) and (\ref{linhom}),
respectively. For higher order corrections, this procedure is repeated
recursively.  It is clear that this expansion holds only provided that
$\fas\ll1$ for all earlier times.

\subsection{How to check the consistency of a power law}

In the previous subsection, we derived an analytic solution for the
density matrix of a single momentum $y$. The
analytical equation for $\eta$ in the power-law region is found by
using the differential equation for $a$ and Eq.~(\ref{etacorr}),
\be \l(\frac{d\eta}{dq}\r)_{\rm analytic} = \int\limits_0^\infty \frac{dy}{2\zeta(3)}
f_{\rm eq}(y) y F l[q,W(q,y,\eta)]~, \label{dZdqanal}\ee
where we have neglected the anti-neutrinos and have used the
conservation of leptonic charge, Eq.~(\ref{conslc}).  We can insert
the solution of $l$ as given in the previous subsection by using the
relation $\tau=q/y$. However, note that the resulting equation for
$\eta$ is valid only as long as $\fas \ll 1$.

Furthermore, we need to define $W$, which is the only parameter in the
analytic solution of $l$ that is dependent on $\eta$ and $y$.
If we define $\eta(q)=\eta_0 q^{-\alpha/3}$, where $\alpha$ is the
temperature behavior of $\eta$, $\eta\propto T^\alpha$, we can write
\be W(\tau,y)=U(\tau)-V_0\eta_0\tau^{-(4+\alpha)/3}y^{-(1+\alpha)/3}~. \label{smartW}\ee
Having given $W$ in this form, we now have two unknown parameters: the
power-law behavior $\alpha$ and the asymmetry factor $\eta_0$.

Checking the consistency of a power law is now simple in principle:
given the correct values for $\alpha$ and $\eta_0$, the integration of
the rhs of (\ref{dZdqanal}) for any $q$ in the power-law regime should yield 
\be \l(\frac{d\eta}{dq}\r)_{\rm analytic}=-\frac{\alpha}{3}~\eta_0
~q^{-1-\alpha/3}~. \label{consrel}\ee

\subsection{Results}
\label{refapp}

We now consider the case derived by \cite{dolgov99},
i.e.~$\alpha=-1$. In this case, (\ref{smartW}) does not depend on $y$
directly, while the $y$-dependence persists indirectly through
$\tau=q/y$. The remarkable consequence is that $\tau_{\rm res}$ is
independent on $y$, and thus is the ratio of $F$ to $\gamma$ at
resonance, $R\equiv F/\gamma_{\rm res}$.

We found it convenient to substitute the integration over $y$ in
Eq.~(\ref{dZdqanal}) with an integration over $\tau$:
\be \frac{d\eta}{dq}= \frac{F q^2}{2\zeta(3)}  \int\limits_0^\infty
\frac{d\tau}{\tau^3}f_{\rm eq}(q/\tau)
l(\tau)~. \label{substdZdqanal}\ee
Also, we use the parameter $R$ instead of $\eta_0$. The advantage of
this is obvious: the $T^{-1}$ power law in
\cite{dolgov99} originates in the counter-term compensating the leading
order term, which occurred when $R={\rm const}={\mathcal O}(1)$. This
yields $\tau_{\rm res}=\sqrt{F/(R\delta)}$. Furthermore, by solving
$W(\tau_{\rm res})=0$, we get
\be |\eta_0(R)|= Q V_0^{-1} \tau_{\rm res}(R) \l[\tau_{\rm res}^{-2}(R)-1\r]~.\label{eta01}\ee

\begin{figure}
\unitlength1mm
\begin{picture}(121,80)
  \put(20,0){\psfig{file=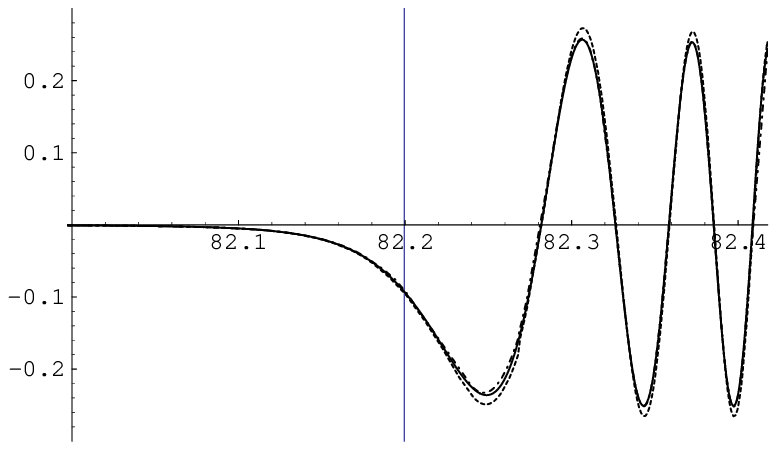,width=10.4cm}} 
  \put(31,68){$l$}
  \put(127,29){$\tau$}
  \put(75,65){$\tau_{\rm res}$}
\end{picture}
  \caption{\label{l-evol} $l(\tau)$ for the case $R=100$,
  $\alpha=-1$. We have plotted the numerical solution (solid line), the
  analytic solution with $f_{(a-s)}=1$ (dashed), and the analytic
  solution with first-order correction (dashed-dotted).}
\unitlength1mm
\begin{picture}(121,80)
  \put(20,0){\psfig{file=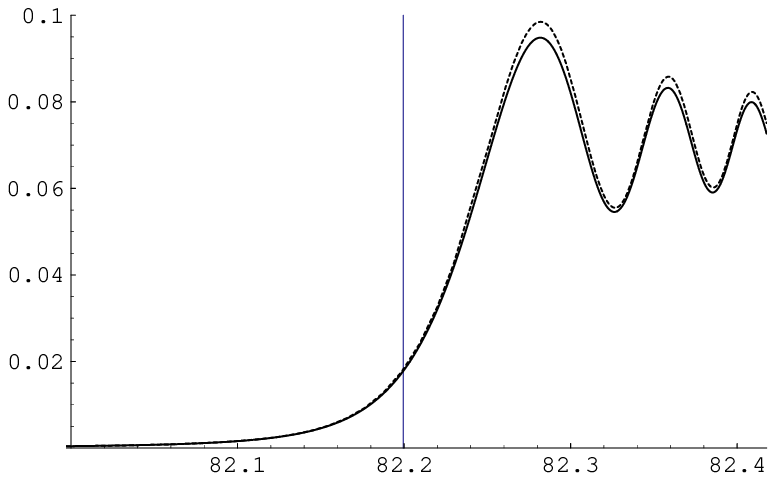,width=10.4cm}} 
  \put(31,68){$s$}
  \put(127,3){$\tau$}
  \put(75,65){$\tau_{\rm res}$}
\end{picture}
  \caption{\label{s-evol} $s(\tau)$ for the case $R=100$,
  $\alpha=-1$. We have plotted the numerical solution (solid line) and the
  analytic solution with $f_{(a-s)}=1$ (dashed).}
\end{figure}

The following calculations were done using the fits presented in
Appendix.
To begin, we have calculated the evolution of $l_{\rm analytic}$ using
$\fas=0$ for $0.1<R<1000$. We have given an example for $R=100$ in
Fig.~\ref{l-evol}. Within this range, $l$ showed a similar behavior,
the only differences occurring in the scaling of $\tau$ and the
amplitude of $l$. Then we calculated the first-order contribution from
$\fas$. In terms of the formalism used in \cite{dolgov99}, this
corresponds to including contributions from $b_0'$, i.e.~the origin of
the counter-term they found. We found that the correction was
negligible for $R<10$, and that it became of order of the main term
only when $R\approx 1000$, and also then stayed smaller than the LO
term.

Thus, we conclude that the counter-term found in \cite{dolgov99}
must be an artifact of the expansion in $F$, which is only valid for
$R\ll1$.

\begin{figure}
\unitlength1mm
\begin{picture}(121,80)
  \put(20,0){\psfig{file=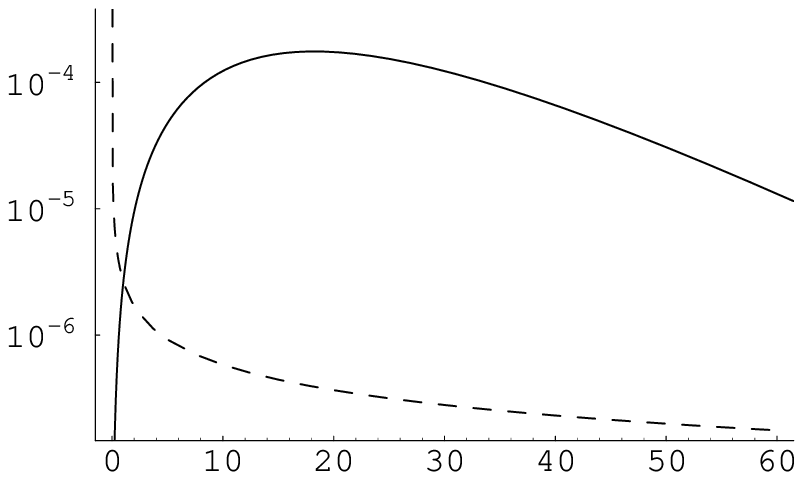,width=10.4cm}} 
  \put(23,68){\large$\frac{d\eta}{dq}$}
  \put(127,4){\large$q$}
\end{picture}
  \caption{\label{comparedetadqs} $\l(\frac{d\eta}{dq}\r)_{\rm analytic}$ (solid)
  and the expected value $\eta_0 q^{-2/3}/3$ (dashed) for the case $R=1$, $\alpha=-1$.}
\end{figure}

Looking at Fig.~\ref{l-evol}, one might argue that even without the
NLO term, the oscillations after the resonance might be the origin of
a large counter-term. However, we found this not to be the case. We
calculated Eq.~(\ref{substdZdqanal}) for $0.2<R<10$. According to
(\ref{consrel}), the result should be equal to $\eta_0 q^{-2/3}/3$ for
all $q$, with $\eta_0$ as given in (\ref{eta01}). We found that this
was not the case, even when including the first-order contribution
from $\fas$: as an example, we show $\eta_0 q^{-2/3}/3$ and
(\ref{substdZdqanal}) in Fig.~\ref{comparedetadqs}, with $R=1$.

By the way, we found that $\l(\frac{d\eta}{dq}\r)_{\rm analytic}$ could be
fitted very well by the simple function
\be \l(\frac{d\eta}{dq}\r)_{\rm analytic} = 5.4 \times10^{-6}\frac{q^2}{R}
f_{\rm eq}\l(\frac{q}{\tau_{\rm res}}\r)~, \ee
in the region $0.2<R<10$, $0.1<q<60$. This corresponds to our
understanding that only the momenta close to resonance are
significant. From this fit, we see that $\l(\frac{d\eta}{dq}\r)_{\rm
analytic}$ has a totally different $q$ behavior than the expected $q^{-2/3}$.

The reason that the oscillations do not bring a counter-term
can be seen in the evolution of $s$, shown in Fig.~\ref{s-evol}. The better part
of the asymmetry is created in the region around the resonance, where
$l$ varies slowly. The oscillations after this region very soon become
fast, and each half period approximately compensates the previous one.


For the case $\alpha\neq-1$, the situation is more complicated, since
then the evolution of $l$ depends on $y$. We intend to eventually present
the results for this case in a later publication.

\begin{figure}
\unitlength1mm
\begin{picture}(121,80)
  \put(20,0){\psfig{file=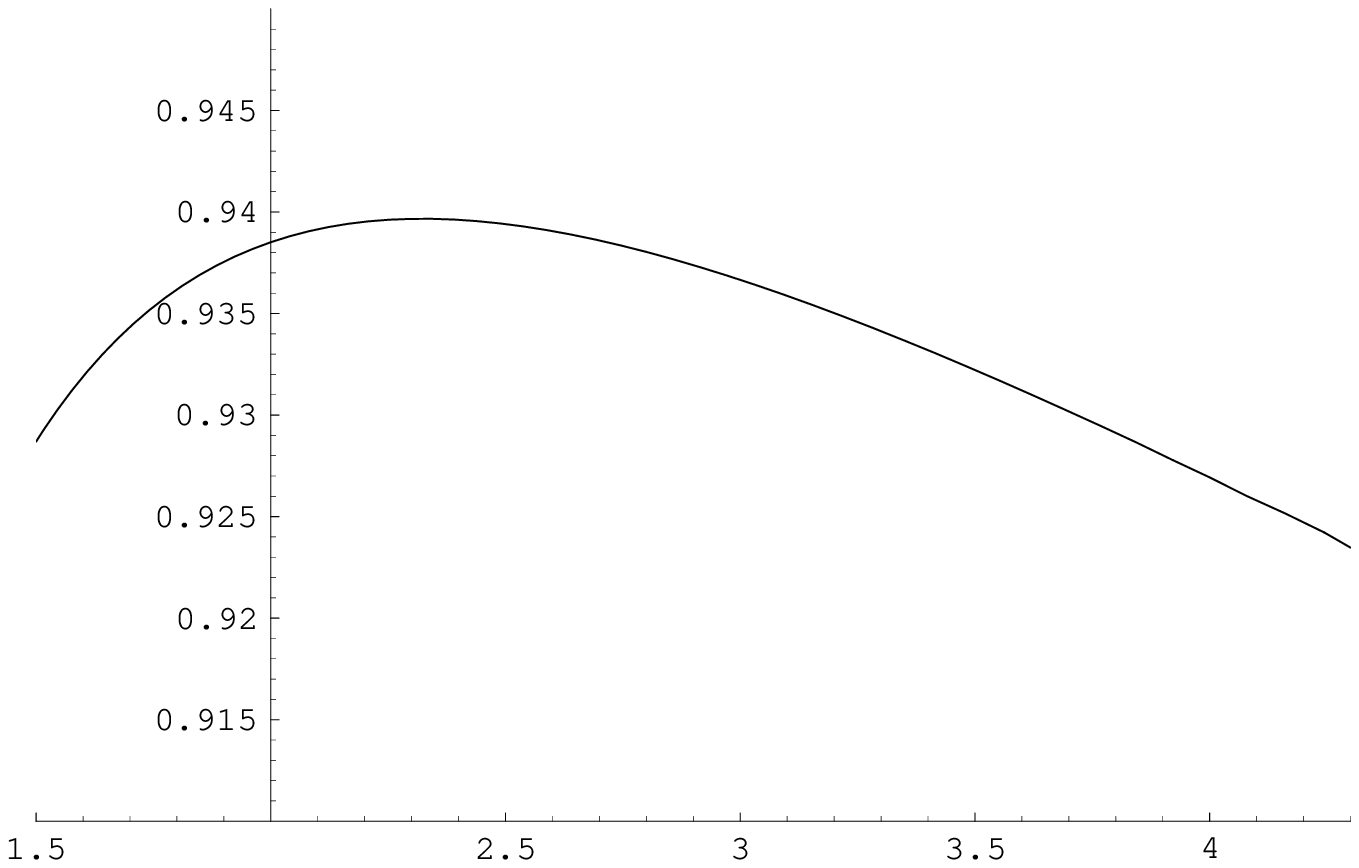,width=10.4cm}} 
  \put(23,68){$\l(\frac{d\eta}{dq}\r)_{\rm num}\big/\l(\frac{d\eta}{dq}\r)_{\rm
expected}$}
  \put(127,2){$q$}
\end{picture}
  \caption{\label{numcompdetadqs}
$\l(\frac{d\eta}{dq}\r)_{\rm num}\big/\l(\frac{d\eta}{dq}\r)_{\rm
expected}$ as a function of $q$ for $\alpha=-11/3$. For simplicity, we
have set $f_{\rm eq}(y_{\rm res})=1/2$, which is a good approximation
for not too large $q$.}
\end{figure}

However, we decided to consider here the special case $\alpha=-11/3$,
i.e.~the power law derived in Eq.~(\ref{derive119}), numerically. To
this end, we assumed this power law, with $\eta_0=(9/11
C_\eta)^{1/3}$, where $C_\eta$ was defined in Eq.~(\ref{defCeta}).
Then we calculated the evolution of $l$ for a grid of momenta $y$
numerically, using the exact differential equations
(\ref{firstasyml}), and integrated the results for several $q$
according to Eq.~(\ref{dZdqanal}). Then we compared the result with
the expected value, in this case given by
\be \l(\frac{d\eta}{dq}\r)_{\rm expected}=\frac{11}{9}\eta_0 q^{2/9}~. \ee
We have plotted the ratio in Fig.~\ref{numcompdetadqs}. If our result
was perfect, the ratio would have been $1$ for all $q$. The
discrepancy has several origins: the result depends on $\alpha$ and
$\eta_0$. As an example, choosing a different $\eta_0$ will shift the
curve upward or downward. Hereby, already a change of a few percent
would suffice to make the curve in Fig.~\ref{numcompdetadqs} cross
unity. Furthermore, a change in $\alpha$ alters the curvature of the
plot. In particular, the downward slope at small $q$ is due to the
fact that we have used the approximation $C(q)\gg q$ when deriving the
power law, see Eqs.~(\ref{detadqint}--\ref{derive119}). In this
context, the result presented in Fig.~\ref{numcompdetadqs} is fairly
good. We conclude that it is safe to say that the real power law is
around $\alpha\approx -4$.

\section{Conclusions}

In this work we have analytically investigated the system of kinetic
equations which governs the creation of lepton asymmetry in the Early
Universe due to active-sterile neutrino oscillations with small vacuum
mixing angle $\sin 2 \theta < 10^{-2}$ and negative $\delta m^2$. We
have improved the analytical approach of \cite{dolgov99}, taking into
account the non-zero chemical potential of neutrinos and relaxing the
adiabatic approximations made in \cite{dolgov99}.

In analogy to \cite{dolgov99}, we have derived an equation for the
asymmetry evolution [Eq.~(\ref{detadq})], which we found to be valid
only if the collision terms dominate the kinetic equations
(\ref{firstasyml}), or in our notation if $F< \gamma$. Except for very
small values of the asymmetry, our approach gave the same results as
found in \cite{dolgov99}. Also, the main term in Eq.~(\ref{detadq})
coincides with the so-called static limit derived in \cite{foot97}.
However, Eq.~(\ref{detadq}) contains an additional term which
compensates the main term when $F\sim \gamma$.

It was found by the authors of \cite{dolgov99} that this additional
term resulted in a growth of the asymmetry according to a $T^{-1}$
power law. However, we found that the additional term in
Eq.~(\ref{detadq}) becomes important only when the expansion breaks
down, or $F/\gamma={\mathcal O}(1)$. But strictly speaking, this fact
does not prove that $\eta\propto T^{-1}$ is wrong, because the
additional term already exists when the perturbative expansion is
valid.

In order to check the validity of the $T^{-1}$ power law, we developed
a new analytical approach which is independent from perturbations in
$F/\gamma$.  We expand the kinetic equations not in the coefficients
(as done in \cite{dolgov99}), but by expanding in the combination of
functions $(a-s-1)$, which remains small as long as the number density
of the sterile neutrinos remains small for all momenta.

This condition strongly depends on the power-law index. For
the $ T^{-1}$ power law, the number density of the sterile neutrinos
remains small for all momenta. This fact allowed us to check the
consistency of $\eta\propto T^{-1}$, and we found that this power law
is not a solution of the kinetic equations. Thus, we proved analytically
that $\eta\propto T^{-1}$ is erroneous.

For the $T^{-4}$ power law, which corresponds to a full transition of
active to sterile neutrinos, the condition $(a-s)\sim 1$ breaks down
already in the region $F/\gamma < 1$.  For the $T^{-11/3}$ power law,
this condition holds until $F/\gamma \sim 1$. Thus, our new method is
not useful beyond this point.

We conclude that our Eq.~(\ref{detadq}) provides a correct analytical
solution, which is in qualitative agreement with the numerical results
of \cite{fv96,footAP99,bfoot00} in the region prior to the formation
of a power law. For $\sin 2 \theta < 5 \times 10^{-4}$, our solution
shows the formation of the power law $ T^{-11/3}$.  On the other hand,
we can not apply our Eq.~(\ref{detadq}) in the region where the MSW
effect dominates and can not decide if this power law changes to $
T^{-4}$ or not. Either way, the final value of asymmetry should be
large.

When this work was nearly complete we received a manucript from P.Di~
Bari, R.~Foot, R.R.~Volkas and Y.Y.Y.~Wong \cite{bfvw00}, who also
conclude that the power law $\eta\propto T^{-1}$ is erroneous by
showing that the approach of \cite{dolgov99} does not work in the
region where the MSW effect dominates over the second-order effects.

\section*{Acknowledgment}
We are grateful to S.~Hansen and G.~Raffelt for helpful comments and
reading the manuscript, and would like to thank M.~Gorbahn,
F.~Gubarev, A.~Neronov and B.~P\"otter for fruitful discussions.

We also like to thank P.~Di Bari, R.~Foot, R.R.~Volkas and Y.Y.Y.~Wong
for making their manuscript \cite{bfvw00} available before publication.
We thank R.~Foot for comments on our manuscript.

This work was partly supported by the Deut\-sche
For\-schungs\-ge\-mein\-schaft under grant No.\ SFB 375 and
in part by INTAS grant 1A-1065.


\begin{appendix}

\section*{Appendix: Fits for {\it l}}

Since the exact analytic function for $l$ is an integral, and we are
forced to integrate over $l$, we have fitted $l$ to speed up the
numerics. Here we present the fits used in our calculations in
subsection \ref{refapp}.

At very small $\tau$, the adiabatic solution, given by $h'=l'=0$, is a
good approximation. However, for small $\gamma$, the condition $h_{\rm
adiab}'\ll W l_{\rm adiab}, \gamma h_{\rm adiab}$ breaks down long
before the resonance. We therefore relax the condition $h'=0$, while
we maintain $l'=0$ (which is valid much longer due to the term $F/2$).
Thus, by setting $l''=l'=0$ in (\ref{delex}), we find
\be l_{\rm semiad} = \frac{F}{2} \frac{\gamma W - W'}{W^3+W \gamma^2 +
W\gamma'-\gamma W'}~. \label{fitsemad} \ee
The error of this solution is of order $l_{\rm semiad}'/(F/2)$.

For the region before and during the resonance, we decided to
calculate the analytic solution of $l$ for a sample of points and fit
a function to it. We found that before the resonance, $l$ showed an
exponential-like behavior, which is why in this region we used a fit
of the form
\be l_{\rm expfit}=e^{\alpha(\tau)}, ~~~~\mbox{where}~~~~
\alpha(\tau)=\alpha_0+\alpha_1 \tau + \dots \ee
Using $n=30$ points and a polynomial of order $(n-1)$, we found it
necessary to do two such fits to achieve an accuracy of better
than $1\%$.

Around the resonance and shortly after, we applied two linear fits,
also with $n=30$.

Finally, we fitted the oscillating behavior after the resonance.
We found that it is a good approximation to add the homogeneous solution
(\ref{lhom}) to the semi-adiabatic fit (\ref{fitsemad}),
\be l_{\rm oscfit}(\tau) = l_{\rm hom}(\tau) + l_{\rm semiad}(\tau)~. \ee
Then the semi-adiabatic fit accounts for the inhomogeneous term

When using the analytic solution for $\tau>\tau_0$, we found that the
integrand diverges at the resonance, $\tau_1=\tau_0$. However, this
problem can easily be circumvented: between $\tau_0$ and some point
$\tau_\epsilon\simgt\tau_0$ we used a quadratic fit for $l$ with
boundary conditions given by the analytic solution at $\tau_i=\tau_0$,
while for $\tau>\tau_\epsilon$ we used the analytic solution with
boundary conditions given by the quadratic fit at
$\tau_i=\tau_\epsilon$. We took $\tau_\epsilon=\tau[W(\tau)=F]$.

We should also mention that for $\tau\ll\tau_0$ the integrand of the
analytic solution is oscillating fast, so that it is difficult to
calculate the integral starting from $\tau=0$. Therefore, we have used
the semi-adiabatic solution up to some $\tau_i\ll\tau_0$, where the
semi-adiabatic solution still is a very good fit, e.g.~has an error of
less than $10^{-3}$.

\end{appendix}

%
%
\nc{\advp}[3]{{\it  Adv.\ in\ Phys.\ }{{\bf #1} {(#2)} {#3}}}
\nc{\annp}[3]{{\it  Ann.\ Phys.\ (N.Y.)\ }{{\bf #1} {(#2)} {#3}}}
\nc{\apl}[3] {{\it  Appl. Phys. Lett. }{{\bf #1} {(#2)} {#3}}}
\nc{\apj}[3] {{\it  Ap.\ J.\ }{{\bf #1} {(#2)} {#3}}}
\nc{\apjl}[3]{{\it  Ap.\ J.\ Lett.\ }{{\bf #1} {(#2)} {#3}}}
\nc{\app}[3] {{\it  Astropart.\ Phys.\ }{{\bf #1} {(#2)} {#3}}}
\nc{\cmp}[3] {{\it  Comm.\ Math.\ Phys.\ }{{ \bf #1} {(#2)} {#3}}}
\nc{\cqg}[3] {{\it  Class.\ Quant.\ Grav.\ }{{\bf #1} {(#2)} {#3}}}
\nc{\epl}[3] {{\it  Europhys.\ Lett.\ }{{\bf #1} {(#2)} {#3}}}
\nc{\ijmp}[3]{{\it  Int.\ J.\ Mod.\ Phys.\ }{{\bf #1} {(#2)} {#3}}}
\nc{\ijtp}[3]{{\it  Int.\ J.\ Theor.\ Phys.\ }{{\bf #1} {(#2)} {#3}}}
\nc{\jmp}[3] {{\it  J.\ Math.\ Phys.\ }{{ \bf #1} {(#2)} {#3}}}
\nc{\jpa}[3] {{\it  J.\ Phys.\ A\ }{{\bf #1} {(#2)} {#3}}}
\nc{\jpc}[3] {{\it  J.\ Phys.\ C\ }{{\bf #1} {(#2)} {#3}}}
\nc{\jap}[3] {{\it  J.\ Appl.\ Phys.\ }{{\bf #1} {(#2)} {#3}}}
\nc{\jpsj}[3]{{\it  J.\ Phys.\ Soc.\ Japan\ }{{\bf #1} {(#2)} {#3}}}
\nc{\lmp}[3] {{\it  Lett.\ Math.\ Phys.\ }{{\bf #1} {(#2)} {#3}}}
\nc{\mpl}[3] {{\it  Mod.\ Phys.\ Lett.\ }{{\bf #1} {(#2)} {#3}}}
\nc{\ncim}[3]{{\it  Nuov.\ Cim.\ }{{\bf #1} {(#2)} {#3}}}
\nc{\np}[3]  {{\it  Nucl.\ Phys.\ }{{\bf #1} {(#2)} {#3}}}
\nc{\pr}[3]  {{\it  Phys.\ Rev.\ }{{\bf #1} {(#2)} {#3}}}
\nc{\pra}[3] {{\it  Phys.\ Rev.\ A\ }{{\bf #1} {(#2)} {#3}}}
\nc{\prb}[3] {{\it  Phys.\ Rev.\ B\ }{{{\bf #1} {(#2)} {#3}}}}
\nc{\prc}[3] {{\it  Phys.\ Rev.\ C\ }{{\bf #1} {(#2)} {#3}}}
\nc{\prd}[3] {{\it  Phys.\ Rev.\ D\ }{{\bf #1} {(#2)} {#3}}}
\nc{\prl}[3] {{\it  Phys.\ Rev.\ Lett.\ }{{\bf #1} {(#2)} {#3}}}
\nc{\pl}[3]  {{\it  Phys.\ Lett.\ }{{\bf #1} {(#2)} {#3}}}
\nc{\prep}[3]{{\it  Phys.\ Rep.\ }{{\bf #1} {(#2)} {#3}}}
\nc{\prsl}[3]{{\it  Proc.\ R.\ Soc.\ London\ }{{\bf #1} {(#2)} {#3}}}
\nc{\ptp}[3] {{\it  Prog.\ Theor.\ Phys.\ }{{\bf #1} {(#2)} {#3}}}
\nc{\ptps}[3]{{\it  Prog\ Theor.\ Phys.\ suppl.\ }{{\bf #1} {(#2)} {#3}}}
\nc{\physa}[3]{{\it Physica\ A\ }{{\bf #1} {(#2)} {#3}}}
\nc{\physb}[3]{{\it Physica\ B\ }{{\bf #1} {(#2)} {#3}}}
\nc{\phys}[3]{{\it  Physica\ }{{\bf #1} {(#2)} {#3}}}
\nc{\rmp}[3] {{\it  Rev.\ Mod.\ Phys.\ }{{\bf #1} {(#2)} {#3}}}
\nc{\rpp}[3] {{\it  Rep.\ Prog.\ Phys.\ }{{\bf #1} {(#2)} {#3}}}
\nc{\sjnp}[3]{{\it  Sov.\ J.\ Nucl.\ Phys.\ }{{\bf #1} {(#2)} {#3}}}
\nc{\sjp}[3] {{\it  Sov.\ J.\ Phys.\ }{{\bf #1} {(#2)} {#3}}}
\nc{\spjetp}[3]{{\it Sov.\ Phys.\ JETP\ }{{\bf #1} {(#2)} {#3}}}
\nc{\yf}[3]  {{\it  Yad.\ Fiz.\ }{{\bf #1} {(#2)} {#3}}}
\nc{\zetp}[3]{{\it  Zh.\ Eksp.\ Teor.\ Fiz.\ }{{\bf #1} {(#2)} {#3}}}
\nc{\zp}[3]  {{\it  Z.\ Phys.\ }{{\bf #1} {(#2)} {#3}}}
\nc{\ibid}[3]{{\sl  ibid.\ }{{\bf #1} {(#2)} {#3}}}

\end{document}